# An extragalactic supernebula confined by gravity

Turner, J. L.,* Beck, S. C.,† Crosthwaite, L. P. ,*‡ Larkin, J. E. ,* McLean, I. S.,* & Meier, D. S. *§

**Department of Physics and Astronomy, UCLA, Los Angeles CA 90049-1562 USA*

*†Department of Physics and Astronomy, University of Tel Aviv, Ramat Aviv, Israel*

*‡Astute Networks, 16868 Via Del Campo Court,Suite 200, San Diego CA 92127*

*§Department of Astronomy, 1002 West Green Street, University of Illinois, Champaign, IL 61801  USA*

**Little is known about the origins of the giant star clusters known as globular clusters. How can hundreds of thousands of stars form simultaneously in a volume only a few light years across—the distance of the sun to its nearest neighbor? Radiation pressure and winds from luminous young stars should disperse the star-forming gas and disrupt the formation of the cluster. Globular clusters in our Galaxy cannot provide answers; they are billions of years old. Here we report the measurement of infrared hydrogen recombination lines from a young, forming super star cluster in the dwarf galaxy, NGC 5253. The lines arise in gas heated by a cluster of an estimated million stars, so young that it is still enshrouded in gas and dust, hidden from optical view. We verify that the cluster contains 4000-6000 massive, hot "O" stars. Our discovery that the gases within the cluster are bound by gravity may explain why these windy and luminous O stars have not yet blown away the gases to allow the cluster to emerge from its birth cocoon. Young clusters**

**in "starbursting" galaxies in the local and distant universe may be similarly gravitationally confined and cloaked from view.**

NGC 5253 is host to hundreds of large star clusters [1], including dozens of extremely bright, young super star clusters [2] (SSCs). Only a few million years old [3,4], these clusters are found in the central "starburst" [4] region of the nearby (3.8 Mpc) galaxy. Subarcsecond radio and infrared imaging reveal [5,6,7] a bright "supernebula" within the starburst, optically invisible, probably powered by a young super star cluster.[8,9] The luminosity required to ionize the supernebula is 0.8-1.2 x $10^9$ $L_\odot$ within a 1-2 pc [8] region, perhaps the most concentrated star-forming luminosity known.

To verify the nature of the supernebula and to study its dynamics, we observed the Brackett α and γ recombination lines of hydrogen at 4.05 μm and 2.17 μm using the NIRSPEC[10] on the Keck Telescope on 11 March 2000. Spectra were taken through a 0.579" x 24" slit at a spectral resolution of R~25,000, or ~12 km/s. SCAM, a 256x256 array camera within NIRSPEC, simultaneously imaged the slit location on the galaxy at K-band (2.2 μm), allowing us to pinpoint where the spectra were taken. Seeing was 0.55"-0.8".

The 2.2 μm broadband image of NGC 5253 reveals hundreds of bright infrared star clusters, shown in Figures 1 and 2. From their optical colors, Calzetti et al.[3] and Tremonti et al.[4] estimate that these SSCs are only ~2-50 Myr in age. We find that the brightest infrared source does not coincide with any of the optical clusters. The brightest 2.2 μm source is offset by ~0.3" ± 0.1" to the northwest of the youngest[3] optical source. This visually-obscured K band source is the location of the strong Brackett line emission that we observe from the supernebula.

Figure 2 shows the image of the slit position with the strongest Brackett line emission and the corresponding Brackett α echellogram. Line plots of both Brackett lines are shown in Figure 3. Continuum-subtracted line fluxes are $S_{Br\alpha}^{obs} = 3.4 \pm 1 \times 10^{-16}$ W m$^{-2}$ and $S_{Br\gamma}^{obs} = 3.9 \pm 1 \times 10^{-17}$ W m$^{-2}$ for a region within ~3" of the continuum peak. For comparison, Kawara, Nishida & Phillips [11] measure $S_{Br\alpha}^{obs} = 7.0 \pm 1 \times 10^{-16}$ W m$^{-2}$ and $S_{Br\gamma}^{obs} = 1.5 \pm 1 \times 10^{-16}$ W m$^{-2}$ in a 10" x 20" aperture; as much as half of their Brackett α flux may be contributed by an unsubtracted continuum. The large aperture flux at K band appears to be dominated by stellar emission, while much of the L' flux is from nebular dust. Our photometry and mapping confirm that most, if not all, of the Brackett α and at least 30% of the Brackett γ emission in NGC 5253 emerges from a source coincident with the brightest K band continuum source.

Having both Brackett lines allows us to correct the line fluxes for extinction and obtain a good estimate of the true, unextincted recombination line flux, and thus the total ionizing flux. For an intrinsic $S_{Br\alpha}^{obs} / S_{Br\gamma}^{obs}$ flux ratio of 2.8 [13], a temperature of 12000 K [14], and a Rieke & Lebofsky [15] extinction law, we obtain extinctions of $A_\alpha = 0.8$ magnitudes at 4.05 μm and $A_\gamma = 2.0$ magnitudes at 2.17 μm. The corresponding optical extinction is $A_v = 18$ magnitudes. Optical Balmer recombination lines give $A_v = 3$ [14]. The apparent contradiction between the low optical extinction and higher infrared extinction is resolved if the extinction is high and internal to the nebula. The extinction-corrected Brackett line fluxes are $S_{Br\alpha}^{obs} = 7.1 \times 10^{-16}$ W m$^{-2}$ and $S_{Br\gamma}^{obs} = 2.5 \times 10^{-16}$ W m$^{-2}$ for the inner ~1.5". These fluxes predict a 15 GHz free-free flux of $24 \pm 8$ mJy: the observed 15 GHz flux in this region, corrected for opacity[9], is $24 \pm 3$ mJy.[6] The Lyman continuum rate required to maintain the ionization of the supernebula is $N_{Lyc} = 4 \pm 1 \times 10^{52}$ s$^{-1}$, equivalent to 4000 O7 stars. This agrees with cm-wave and mm-wave free-free fluxes [6,8,9] and radio recombination lines[16] (Owens Valley mm-wave fluxes[9] give a value of 6000 O7 stars from a larger aperture). Both the excellent agreement of the Brackett line fluxes and radio fluxes and the observed compactness of the emission argue that the Brackett line emission arises from the radio "supernebula".

The velocity information presented here is new; these spectra probe the dynamics of the nebula at high spatial and spectral resolution. The Brackett line profiles were fit with Gaussians, with line centroids at $v_{LSR}$ = 377 ± 2 km s$^{-1}$ and full widths half maxima of 76±1 km s$^{-1}$, consistent with H$\alpha$[17] and radio recombination lines[16]. For gas at 12,000 K, this linewidth is supersonic. Supersonic gas motions are expected in nebulae: in addition to nebular expansion there is also the interaction of the nebular gas with winds from the massive stars in the cluster. For comparison, compact HII regions in our Galaxy around small groups of O stars have linewidths of up to 64 km/s[18]; nebulae around individual O stars in the Wolf-Rayet phase can have linewidths of 50-200 km/s. [19]

The first implication of the linewidth of 75 km/s for the "supernebula" is that if the nebula were actually expanding at the implied speed of ~38 km/s, then the mean radius of ~0.7 pc[8] of the nebula implies an implausibly short dynamical age of 7000 years. Dynamical ages of compact HII regions in our Galaxy are generally too short to explain the numbers of observed nebulae [20]. It is thought that confinement mechanisms such as the pressure of dense molecular clouds are responsible for the retardation of their expansion. This could lengthen the lifetime of the supernebula phase, if there were molecular gas nearby, which is as yet undetected[9].

The more unusual implication of the linewidth is that while supersonic, these nebular lines are remarkably narrow given the size of the nebula and the high luminosity of its embedded star cluster. Brackett line, radio, and infrared fluxes all require $L_{OB}$ ~ 0.8-1.2 x 10$^9$ L$_\odot$ for excitation of the nebula. For a cluster with a Salpeter initial mass function (IMF) and a lower mass cutoff of 1 M$_\odot$, the mass in stars corresponding to this luminosity is 5-7 x 10$^5$ M$_\odot$; if the IMF extends down to stars less than 1 M$_\odot$, the cluster mass may reach $10^6$ $M_\odot$. The size of the radio nebula is well-determined, and should be the same as the Brackett size since both emissions are proportional to $\int n_e^2 dl$; VLA images show that the nebula is 0.9 pc x 1.8 pc in size (0.05" x 0.1", ± 0.02") [8]. We assume that the star cluster exciting the nebula lies within the supernebula or else the implied excitation luminosity would be higher than the total observed IRAS luminosity of the entire galaxy, 1.8 x $10^9$ $L_\odot$[7]. For a radius of 0.5-0.9 pc, the escape velocity is ~25-30 km s$^{-1}$ for a cluster of O stars alone; $v_{esc}$ ~50-70 km s$^{-1}$ for an IMF cutoff at 1 $M_\odot$; and $v_{esc}$ ~85-110 km s$^{-1}$ for a cluster IMF extending below 1 $M_\odot$. Gravity must play a significant role in the dynamics of this nebula, slowing its expansion if not halting it. Virial linewidths are only slightly smaller than $v_{esc}$, so the nebula could actually be in gravitational equilibrium.

If the "supernebula" stage of super star cluster formation should prove to be common and long-lived, it may have implications for the detection of Lyman α and other ultraviolet lines in star-forming galaxies. There is evidence that Ly α is detectable primarily in galaxies in which the potential absorbing gas is Doppler-shifted out of the path of the Ly α photons.[21] This would be most likely in galaxies with superwinds and superbubbles, phenomena which can be caused by large concentrations of hot and windy O stars.[22] If a significant fraction of the lifetime of the ionizing phase of a super star cluster is spent in a confined, extincted state such as the supernebula, it is likely that these winds will develop only late in the formation process of the cluster, perhaps at the occurrence of the first supernovae. By that time the ionizing flux of the cluster is already declining. Super star clusters may be an important mode of star formation in the early universe. Ly α searches for primeval galaxies may therefore fall far short of detecting the true star formation rate, as suggested by observations.[21,23,24,25]

To put the supernebula in context we can compare it to a more traditional nebula, 30 Doradus in the Large Magellanic Cloud. 30 Doradus is a large, luminous nebula ionized by an optically visible star cluster with an estimated age of <10 Myr,[26] ** 150-200 pc in extent and with a gas density of 20-100 cm$^{-3}$.[27] 30 Doradus is 200 times larger and 100-1000 times less dense than the supernebula in NGC 5253, and its exciting star cluster R136 is ten to 100 times less massive than the cluster in NGC 5253.[28] The escape velocity for 30 Doradus is less than the thermal sound speed of 10 km/sec so its gas motions are determined by winds and turbulence rather than gravity.[29] How will the supernebula in NGC 5253 free itself from gravitational bondage and evolve into an extended, diffuse nebula like 30 Doradus? That will probably depend on the evolution of its underlying star cluster.

The authors thank Tiffany Glassman and Eric Greisen for assistance with the data and Michael Jura for helpful discussions. This research is supported by NSF grant AST00-71276, and a grant from the Israel Academy Center for Multi-Wavelength Astronomy.

**Correspondence and requests for materials should be addressed to J.T. (e-mail: turner@astro.ucla.edu).**

Figure 1. Infrared-optical color view of the young super star clusters of NGC 5253. The $\lambda$ = 2.2$\mu$m image (red channel) was constructed from SCAM images made during the night. The infrared image is combined with an optical image from the Hubble Space Telescope (blue and green channels). Seeing was ~0.7-0.8" for the infrared image; the HST image was convolved to match. The brightest K-band source appears as an extended red source to the north of the dust lane. A Gaussian fit to this source gives a size (FWHM) of 1.10" x 0.9", p.a. 36°. From smaller clusters we estimate the point spread function for the SCAM image to be 0.75" ± 0.2". If the K-band source is Gaussian, this would imply a source size of ~0.8" x 0.5"; along p.a. 40°. This size is uncertain because of variable seeing. The entire SCAM image is 46" square. The orientation of the image is east to the left, north up.

Figure 2. Echellogram and slit position for Brackett spectra of the supernebula. The 46"x46" SCAM image shows the position of the 0.579" x 24" slit on the brightest K-band source in NGC 5253. In the inset is the Brackett $\alpha$ echellogram, with frequency/velocity running horizontally and the spatial dimension vertically. The full slit length is 24", but only 6" is shown in the inset. The orientation of the image is such that north is at an angle of -43° (clockwise) from vertical. The Brackett line emission is less than 1.3" in spatial extent on the echellogram, the same as the standard star. Spectra taken at eight other positions showed weak or no emission

Figure 3. Spectra of Br $\alpha$ and Br $\gamma$ in NGC 5253. The spectra were integrated over the central 3" centered on the brightest infrared continuum source. Each grating setting was followed by a calibration A star. On/off slit SCAM images of the standard indicate that 50% of the light entered the slit. Since the Brackett line emission has spatial extent comparable to the standard star, ~6-9 pixels (0.8"-1.3"), calibration using the standard should automatically correct for this effect. We estimate an uncertainty of 30% in the line fluxes due to variability in seeing. Near-infrared continuum emission from the brightest K band source is so strong that it can be seen even in these highly dispersed spectra. We measure continuum fluxes of 186±60 mJy and 12 ± 4 mJy at L' and K,

respectively, as compared to the 144 mJy and 20 mJy fluxes at in these bands observed by Moorwood & Glass [12] with 8-9" apertures.

**Note added after publication: The age of the R136 cluster is significantly less than the 10 Myr quoted in the published article, as corrected here. Our thanks to Nolan Walborn for catching this error. Conclusions are unaffected.

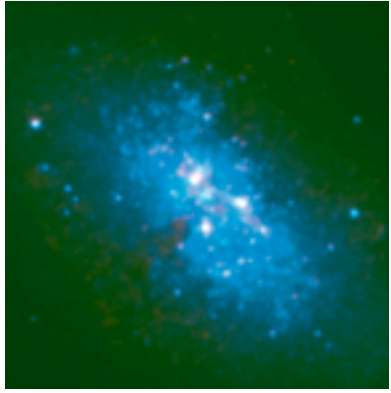

Figure 1

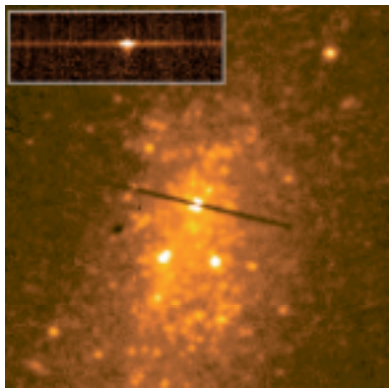

Figure 2

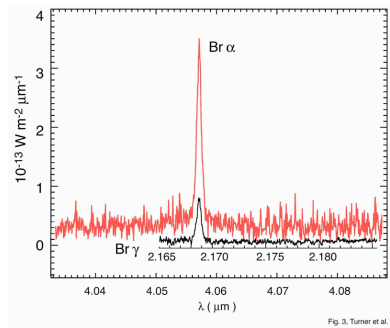